\documentclass[12pt]{article}%
\usepackage[T2A]{fontenc}
\usepackage[cp1251]{inputenc}
\usepackage{amsmath}
\usepackage{amsfonts}
\usepackage{amssymb}
\usepackage{graphicx}
\usepackage{cite}%

\begin{document}

\title{Matched shadow processing}
\author{A.L. Virovlyansky\\
\textit{{\small Institute of Applied Physics, Russian Academy of Science}}\\
\textit{{\small 46 Ul'yanov Street, 603950 Nizhny Novgorod,
Russia}} \\
\textit{{\small virovlyansky@mail.ru}}\\
\date{}} \maketitle

\begin{abstract}
Traditional matched field processing is based on the comparison of the complex
amplitudes of the measured and calculated wave fields at the aperture of the
receiving antenna. This paper considers an alternative approach based on
comparing the intensity distributions of these fields in the "depth -- arrival
angle" plane. To construct these intensities, the formalism of coherent states
borrowed from quantum mechanics is used. The main advantage of the approach
under consideration is its low sensitivity to the inevitable inaccuracy of an
environmental model used in calculation.

\end{abstract}

\newpage

\section{Introduction}

The applicability region of the traditional matched field processing (MFP)
\cite{Bucker76,Hinich73}, based on the comparison of the complex amplitudes of
the measured and calculated sound fields, is substantially limited by the fact
that the complex field amplitude -- especially at high frequencies -- is very
sensitive to variations in the environmental parameters. Therefore, the
inevitable inaccuracies of the mathematical model of the environment used in
calculations often make the application of MFP impossible
\cite{Gerstoft1995,Baggeroer2013,JKPS2011}.

This paper discusses an alternative approach for comparing measured and
calculated fields. Here we use the field expansion in coherent states
\cite{Klauder,V2017}. The intensity (amplitude squared) of the coherent state
can be interpreted as a contribution to the total energy of the sound field
from waves arriving at a given depth interval at grazing angles from a given
angular interval. The set of amplitudes squared of all coherent states gives
the distribution of the sound field intensity in the phase plane "depth --
arrival angle". In quantum mechanics, an analogous characteristic of a wave
function is called the Husimi distribution \cite{Husimi,Vbook2010}.

Our idea is that instead of comparing the complex amplitudes of the measured
and calculated fields, we can compare the intensity distributions of these
fields in the phase space, which are less sensitive to variations in the
environmental parameters. We divide the phase space into two zones: an
insonified zone in which the field intensity is relatively large, and a shadow
zone where the intensity is close to zero. The parameter by which the fields
are compared is the ratio of the integral intensities in the insonified zone
and in the shadow zone. In fact, our method is based on comparing the areas
occupied by the shadow zones of the measured and calculated fields (at the
same time, the areas occupied by the insonified zones are also compared).
Therefore, we call this approach the matched shadow processing (MSP).

The proposed approach is motivated by the results of the recent work
\cite{V2017}, where the expansion in coherent states is used to isolate the
components of the sound field which are low sensitive to sound speed fluctuations.

In Sec. \ref{sec:desitrub}, a definition of the field intensity distribution
in phase plane is given and an example is presented of calculating such a
distribution for an idealized model of a waveguide in a shallow sea. The
similarity coefficient of the measured and calculated fields, based on a
comparison of the distributions of their intensities, is introduced in Sec.
\ref{sec:coeff}. In Secs. \ref{sec:variations} and \ref{sec:localization} it
is shown that this coefficient is much less sensitive to variations in the
environmental parameters and source position than the similarity coefficient
used in the traditional MFP.

\section{Field intensity distribution in the phase plane \label{sec:desitrub}}

Consider a CW sound field excited by a point source at frequency $f$ in
a two-dimensional underwater waveguide with the sound speed field
$c\left(  r,z\right)  $, where $r$ is the range and $z$ is the depth. The
refractive index is $n\left(  r,z\right)  =c_{0}/c(r,z)$, where $c_{0}$ is the
reference sound speed satisfying condition $\left\vert c\left(  r,z\right)
-c_{0}\right\vert \ll c_{0}$.

In the scope of the Hamiltonian formalism the ray path at range $r$ is
determined by its vertical coordinate $z$ and momentum $p=n\left(  r,z\right)
\sin\chi$, where $\chi$ is the ray grazing angle. It is assumed that the sound
field is excited by a point source set at point $\left(  0,z_{s}\right)  $.
Different rays escape this points with different starting momenta $p_{0}$.
Momenta and depths of rays at the observation range $r$ are described by
functions $p\left(  p_{0}\right)  $ and $z\left(  p_{0}\right)  $, respectively.
In phase plane (momentum $P$ -- depth $Z$), the arrivals of rays form a curve,
which we call the ray line. It is defined by the relations $P=p\left(
p_{0}\right)  $~and $Z=z\left(  p_{0}\right)  $.

The ray line establishes a correspondence between the depths and the arrival
angles of the rays. Changing the environmental parameters or the source
position is manifested in the change of this line. Therefore, the comparison
of the ray lines of the measured and calculated fields can be used in solving
inverse problems. Although the uncertainty principle does not allow the
accurate reconstruction of the ray line, one can restore a fuzzy version of
this line. In Ref. \cite{V2017} it is shown that this can be done using the
field expansion in coherent states.

The coherent state associated with the point of the phase plane $\mu=\left(
P,Z\right)  $ is given by the function \cite{Klauder,V2017,LLquant}
\begin{equation}
Y\left(  z;P,Z\right)  =\frac{1}{\sqrt{\Delta_{z}}}\exp\left[  ikP\left(
z-Z\right)  -\frac{\pi\left(  z-Z\right)  ^{2}}{2\Delta_{z}^{2}}\right]  ,
\label{Y-z}%
\end{equation}
where $k = 2\pi f/c_0$, $\Delta_{z}$ is the width of the coherent state along the $z$-axis. It
describes waves arriving at depths close to $Z$ at grazing angles close to
$\chi=\arctan P$. The scalar product of coherent states associated with points
$\mu_{1}=\left(  P_{1},Z_{1}\right)  $ and $\mu_{2}=\left(  P_{2}%
,Z_{2}\right)  $ is%
\begin{equation*}
\left\vert \int dz~Y_{\mu_{1}}\left(  z\right)  Y_{\mu_{2}}^{\ast}\left(
z\right)  \right\vert =\frac{1}{\sqrt{2}\Delta_{z}}\exp\left(  -\frac{\pi}%
{2}d\left(  \mu_{1},\mu_{2}\right)  \right)  ,%
\end{equation*}
where the asterisk denotes complex conjugation,
\begin{equation*}
d\left(  \mu_{1},\mu_{2}\right)  =\frac{\left(  P_{2}-P_{1}\right)  ^{2}%
}{\Delta_{p}^{2}}+\frac{\left(  Z_{2}-Z_{1}\right)  ^{2}}{\Delta_{z}^{2}},
\end{equation*}
$\Delta_{p}=\lambda/(2\Delta_{z})$, $\lambda=2\pi/k$ is the wavelength.
Function $d\left(  \mu_{1},\mu_{2}\right)  $ can be interpreted as a
dimensionless distance between the points of the phase plane $\mu_{1}$ and
$\mu_{2}$. The coherent states associated with these points will be assumed to
be close for $d<1$ and different for $d>1$.

Let us define the amplitude of the coherent state as
\begin{equation}
a_{\mu}=\int dz~u\left(  z\right)  Y_{\mu}^{\ast}\left(  z\right)  .
\label{a-u}%
\end{equation}
The dependence $\left\vert a_{\mu}\right\vert ^{2}$ on $\mu$ will be
interpreted as the distribution of the intensity of the sound field in the
phase plane $P-Z$. In quantum mechanics $\left\vert a_{\mu}\right\vert ^{2}$
is called the Husimi distribution. It is the Wigner function of the wave field
smoothed over the elementary phase plane cell of size $\Delta_{p}\times
\Delta_{z}$ \cite{Vbook2010,Tatar}. \ The distribution of intensity
$\left\vert a_{\mu}\right\vert ^{2}$ is localized inside the region, which is
formed by points located at distances $d<1$ from the ray line (the distance
from a point to a line is the distance to the nearest point of this line). We
will call this area the fuzzy ray line.

As an example, let us consider an idealized model of a waveguide in a shallow
sea. This is a range-independent waveguide of depth $h$ with a linear sound
speed profile. At depths $z=0$ and $z=h$, the sound speed takes values $c_{0}$
and $c$, respectively. The value of $c_{0}$ is taken to be 1.5 km/s.

A waveguide with depth $h=h_{a}$ and $c=c_{a}$, where $h_{a}=$ 190 m and
$c_{a}=$ 1.48 km/s,  is considered as \textit{unperturbed}. The sound speed
profile in this waveguide is shown in the left panel of Fig. 1. The
perturbation is modeled by the deviations of $h$ and $c$ from $h_{a}$ and
$c_{a}$, respectively. In addition, by perturbation we also mean changing the
source position. As unperturbed values of the source depth and its distance to
the antenna, we take $z_{a}=$ 90 m and $r_{a}=$ 2 km, respectively. All
calculations are performed at a carrier frequency $f=$ 2000 Hz. The bottom is
a liquid half-space with a density of 1400 kg /m$^{3}$ and a sound speed of
1.6 km / s.

\begin{figure}[ptb]
\begin{center}
\includegraphics[
height=7.cm, width=15.cm ]{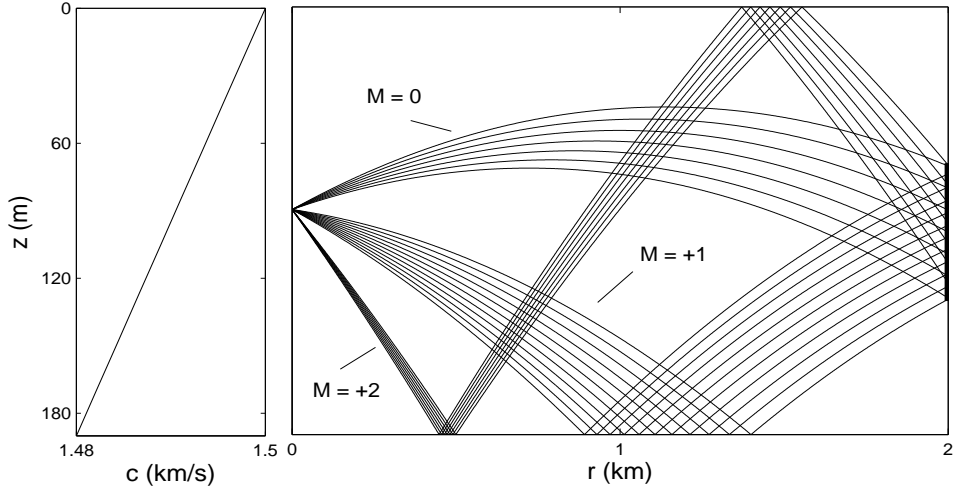}
\end{center}
\caption{Left panel: unperturbed sound speed profile. Right panel: beams of
rays with idenitifiers $M=0$, $M=+1$, and $M=+2$ hitting the antenna.
}%
\label{fig1}%
\end{figure}

Using a mode code, the complex amplitude of the wave field is computed at the
aperture of receiving antenna, covering the depth interval from $z_{1}=$ 50 m
to $z_{2}=$ 150 m and located at different ranges $r$ from the source. When
expanding the wave field in coherent states, we take into account only those
states that are formed by waves with grazing angles $\left\vert \chi
\right\vert <18^{\circ}$. Waves with such $\chi$ are formed by modes of the
discrete spectrum. This condition means that we consider the intensity
$\left\vert a_{\mu}\right\vert ^{2}$ at the points of the phase plane with
$\left\vert P\right\vert <\sin18^{\circ}=0.31$. According to Eqs. (\ref{Y-z})
and (\ref{a-u}), to calculate the amplitude of the state associated with point
$\left(  P,Z\right)  $, one should know the field in the depth interval
$Z\pm\Delta_{z}/2$. Therefore, we can find the amplitudes of only those states
that are associated with the points of the phase plane at depths $z_{1}%
+\Delta_{z}/2<Z<z_{2}-\Delta_{z}/2$.

In what follows we consider fields at observation ranges of 1.5 to 2.5 km. At
such distances, the field on the antenna is formed by relatively narrow beams
of rays, examples of which are shown in the right panel Fig. 2. Each beam
includes rays with the same identifier $\pm M$, where $M$ is the number of
reflections from the waveguide boundaries, and the + (-) sign means that after
leaving the source the first time the ray is reflected from the bottom
(surface). At the observation range, the rays of each beam form a segment of
the ray line.

\begin{figure}[ptb]
\begin{center}
\includegraphics[
height=10.cm, width=15.cm ]{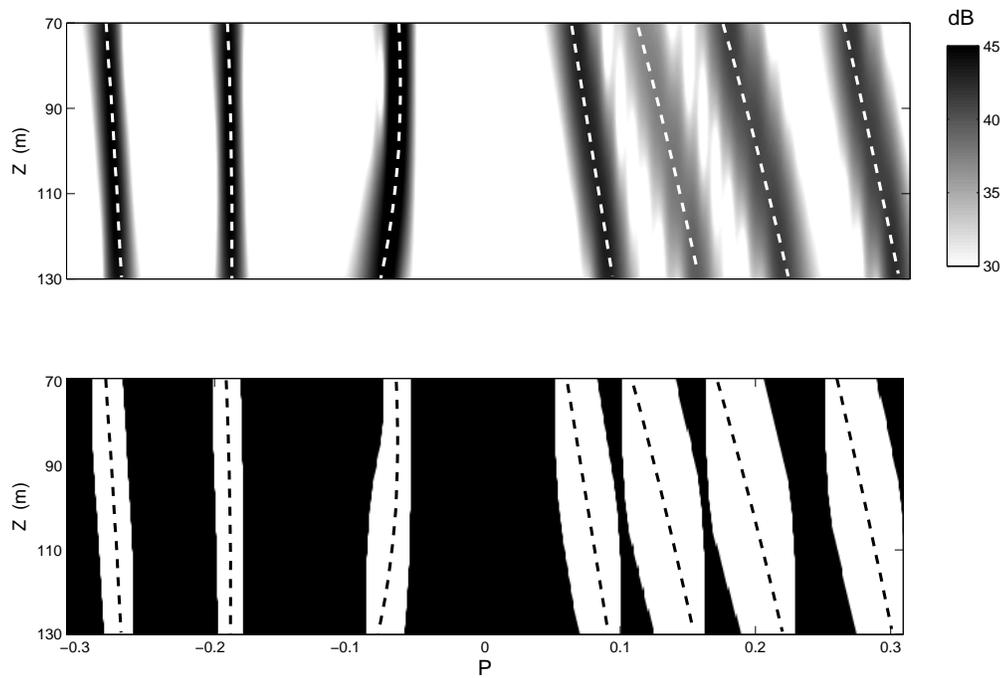}
\end{center}
\caption{Upper panel: distribution of field intensity $|a_{\mu}|^{2}$ in the
phase plane $P-Z$ in the unperturbed waveguide. Lower panel: the insonified
zone and the shadow zone are shown in white and black, respectively. The same
dashed lines in both panels depict the segments of the ray lines formed by
rays with identifiers (from left to right) +3, -2, +1, 0, -1, +2, and -3. }%
\label{fig2}%
\end{figure}

Such segments in the unperturbed waveguide are shown in Fig. 2 by dashed
curves. The upper panel of this figure presents the intensity distribution
$\left\vert a_{\mu}\right\vert ^{2}$ at $r=2$ km. Here and below, the
projections onto coherent states are calculated with $\Delta_{z}=40$ m. This
value of the vertical scale is chosen empirically for a good resolution of
fuzzy segments of the ray line.

White areas in the bottom panel of Fig. 2 are formed by points located at
distances $d<1$ from the segments of the ray line. These areas form a fuzzy
ray line. It represents the insonified zone of the phase plane. We denote this
zone $\sigma_{in}$. The remaining part of the phase plane (it is shown in
black) is the shadow zone. It will be denoted by $\sigma_{sh}$.

\section{Similarity coefficients \label{sec:coeff}}

Let $u\left(  z\right)  $ and $\tilde{u}\left(  z\right)  $ be the complex
amplitudes of the sound fields at the receiving antenna in the unperturbed and
perturbed waveguide, respectively. The intensities distributions of these
fields in the phase plane are $\left\vert a_{\mu}\right\vert ^{2}$ and
$\left\vert \tilde{a}_{\mu}\right\vert ^{2}$. In the traditional MFP, the
similarity of fields $u$ and $\tilde{u}$ is quantitatively characterized by
the coefficient \cite{Bucker76}%

\begin{equation*}
K=\frac{\left\vert \int dz~ u\left(  z\right)  \tilde{u}^{\ast}\left(
z\right)  \right\vert }{\left(  \int dz~\left\vert u\left(  z\right)
\right\vert ^{2}\right)  ^{1/2}\left(  \int dz~\left\vert \tilde{u}\left(
z\right)  \right\vert ^{2}\right)  ^{1/2}}. %
\end{equation*}

The main idea of this work is to introduce another similarity coefficient,
which is much less sensitive to variations in environmental parameters and
source position. This coefficient is $Q=\tilde{q}/q$, where%
\begin{equation*}
q=\left.  \int_{\sigma_{in}}d\mu~\left\vert a_{\mu}\right\vert ^{2}\right/
\int_{\sigma_{sh}}d\mu~\left\vert a_{\mu}\right\vert ^{2},\;\tilde{q}=\left.
\int_{\sigma_{in}}d\mu~\left\vert \tilde{a}_{\mu}\right\vert ^{2}\right/
\int_{\sigma_{sh}}d\mu~\left\vert \tilde{a}_{\mu}\right\vert ^{2}.
\end{equation*}
In these formulas, $q$ is the ratio of the field intensity in the unperturbed
waveguide integrated over the insonified zone $\sigma_{in}$ to the intensity
integrated over the shadow region $\sigma_{sh}$. The quantity $\tilde{q}$
represents the ratio of the field intensities in the perturbed waveguide
integrated over \textit{the same} zones of the phase plane that were found for
the unperturbed waveguide.

Below, with concrete examples, we show that the coefficient $Q$ is indeed much
less sensitive to variations in environmental parameters and source position
than $K$.

\section{Variations in medium parameters and source range
\label{sec:variations}}

In this section, the sound field $u\left(  z\right)  $ on the antenna located
at range $r=r_{a}$ in the unperturbed waveguide (with $h=h_{a}$ and $c=c_{a}$)
is compared with the fields at the same antenna $\tilde{u}\left(  z\right)  $,
calculated for other values of $r$, $h$, and $c$. Calculations performed using
the MFP and MSP methods for a fixed source depth $z_{s}=z_{a}$ allow us to
find the dependences of the similarity coefficients $K$ and $Q$ on $r$, $h$
and $c$.

Figure 3 shows sections of the uncertainty functions $K(r,h,c)$ (a, c, e) and
$Q(r,h,c)$ (b, d, f) by planes $r=r_{a}$ (a,b), $h=h_{a}$ (c,d), and $c=c_{a}$
(e,f). As it is seen, the dependencies of these functions on the arguments
$r$, $h$ and $c$ are radically different. Both take their maximum values
$K=Q=1$ at the point $\left(  r_{a},h_{a},c_{a}\right)  $. However, the
function $Q(r,h,c)$ gradually decreases with deviation from this point. In
contrast, the function $K(r,h,c)$, in addition to the sharp peak at the point
$\left(  r_{a},h_{a},c_{a}\right)  $, has many more local maxima.

\begin{figure}[ptb]
\begin{center}
\includegraphics[
height=12.cm, width=18.cm ]{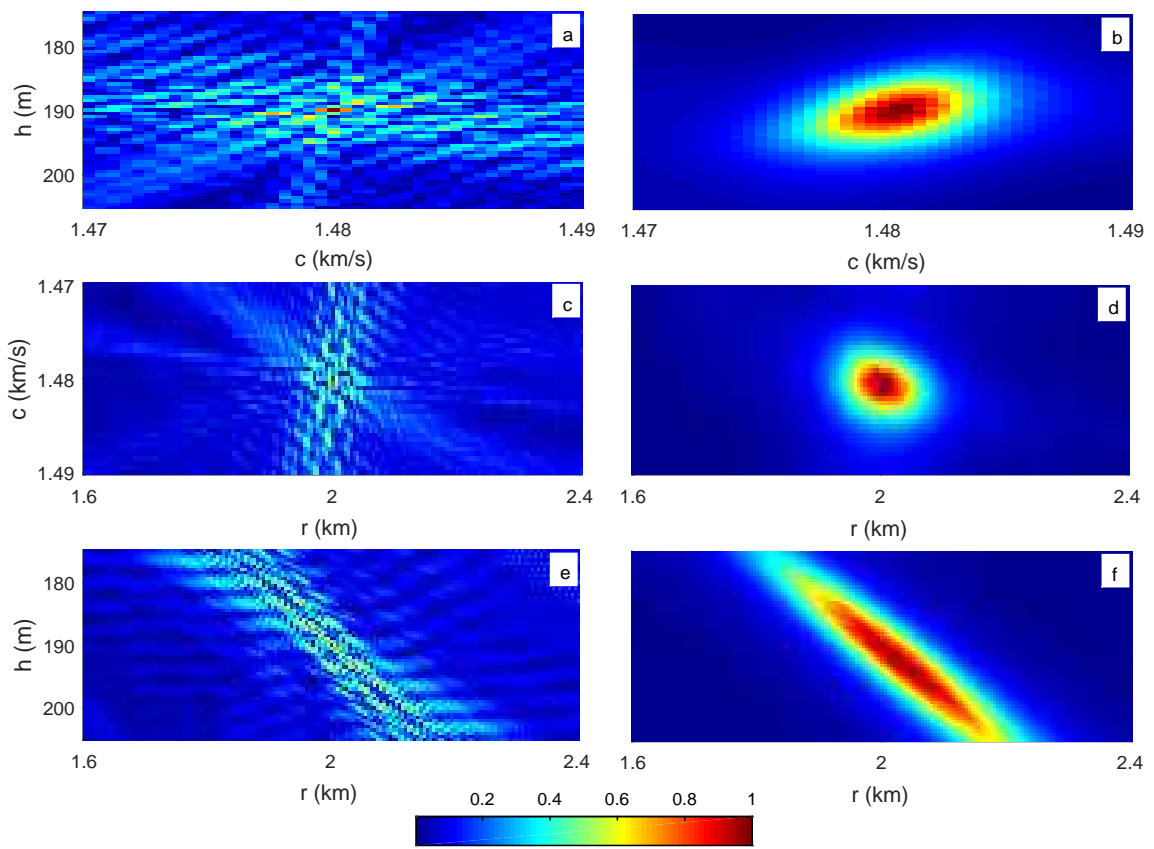}
\end{center}
\caption{Sections of uncertainty functions $K(r,h,c)$ (left column of plots)
and $Q(r,h,c)$ (right column) by planes $r=r_{a}$ (a,b), $h=h_{a}$ (c,d), and
$c=c_{a}$ (e,f).}%
\label{fig3}%
\end{figure}

\section{Variations of source coordinates \label{sec:localization}}

Let us turn to an analysis of the similarity coefficients of the perturbed and
unperturbed fields, which are excited by sources located at different
distances to the antenna $r_{s}$ and at different depths $z_{s}$. Here we
simulate the situation that arises when solving the problem of source
localization. Let us assume that the actual values of the distance from the
source to the antenna and the source depth are $r_{s}=r_{a}$ and $z_{s}=z_{a}%
$. The sound field $\tilde{u}(z)$ at the antenna, calculated for a source with
such coordinates in, generally speaking, a perturbed waveguide, is regarded as
"measured". To restore the source position, the measured field $\tilde{u}(z)$
is compared with the fields $u(z)$ calculated in the unperturbed waveguide for
different values of the source coordinates $z_{s}$ and $r_{s}$. The comparison
results are expressed by the uncertainty functions $K\left(  r_{s}%
,z_{s}\right)  $ and $Q\left(  r_{s},z_{s}\right)  $ representing the
dependencies of the similarity coefficients $K$ and $Q$ of the measured and
calculated fields on the trial source position ($r_{s}$, $z_{s}$).

Figure 4 shows the results of calculating the functions $K\left(  r_{s}%
,z_{s}\right)  $ (a,c,e) and $Q\left(  r_{s},z_{s}\right)  $ (b,d, f) in the
unperturbed (a,b), slightly perturbed (c,d) and strongly perturbed (e,f)
waveguide. The calculations are performed for $r_{s}$ in the interval from 1.5
km to 2.5 km and $z_{s}$ from 20 to 150 m. The $h$ and $c$ parameters of the
waveguides used in the simulation are indicated in the plots.

In the unperturbed waveguide, both the functions $K\left(  r_{s},z_{s}\right)
$ (a) and $Q\left(  r_{s},z_{s}\right)  $ (b) have peaks with centers at the
point $\left(  r_{a},z_{a}\right)  $. The peak of function $K\left(
r_{s},z_{s}\right)  $ is very narrow, and in perturbed waveguides (c,e) it splits into  sets of local maxima. In contrast, the function $Q\left(  r_{s},z_{s}\right)  $ has a much wider peak, which, however, remains in the perturbed waveguides (d,f).

\begin{figure}[ptb]
\begin{center}
\includegraphics[
height=12.cm, width=18.cm ]{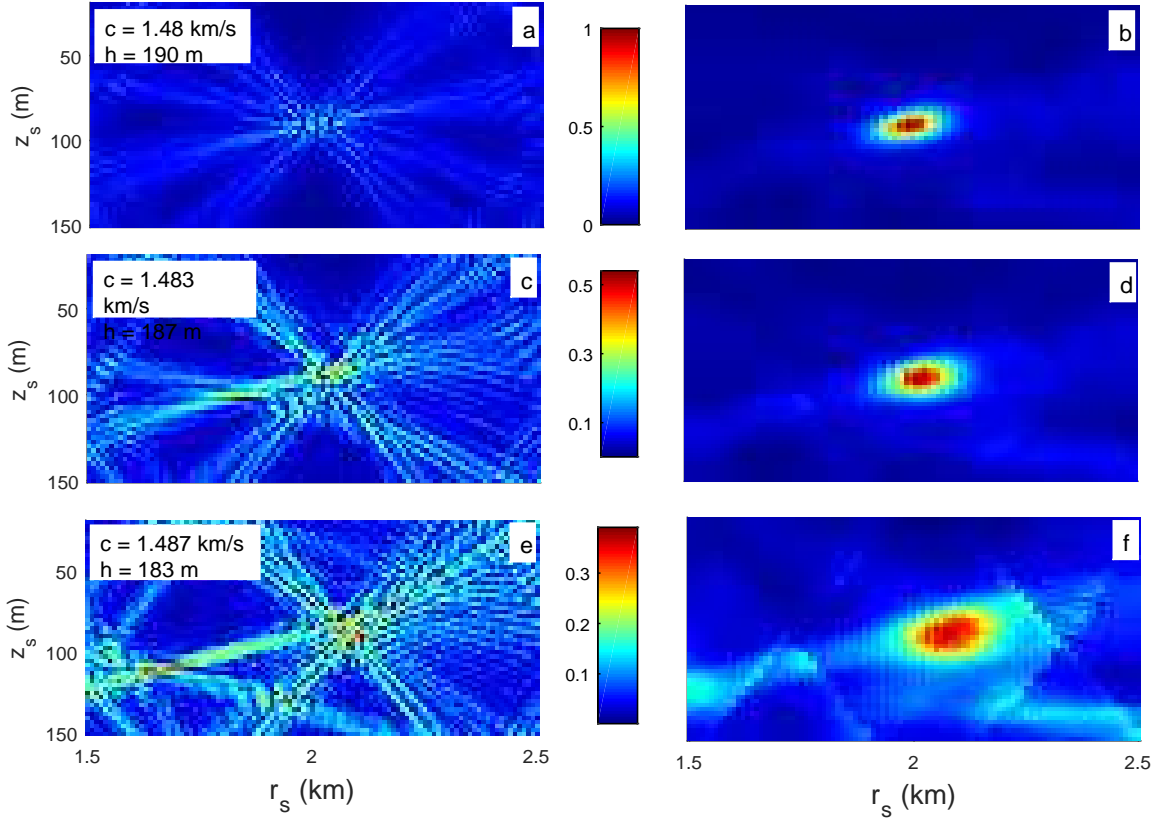}
\end{center}
\caption{Uncertainty functions $K(r_{s},z_{s})$ (left column of plots) and
$Q(r_{s},z_{s})$ (right column) in the unperturbed (a,b), weakly perturbed
(c,d), and strongly perturbed waveguide (e,f).}%
\label{fig4}%
\end{figure}

The abrupt changes in the coefficient $Q$ in Figs. 3b, d, and f, which are
observed near some inclined lines, can be explained as follows. The
intersection of these lines by the point $\left(  r_{s},z_{s}\right)  $ causes
a change in the set of identifiers $\pm M$ of rays arriving at the antenna at
grazing angles $\left\vert \chi\right\vert <18^{\circ}$ used in constructing
the zones $\sigma_{in}$ and $\sigma_{sh}$. This leads to a change in these
zones and, accordingly, to a jump in $Q$.

The simulation results suggest that based on MSP, a robust source localization
algorithm can be created that is applicable in conditions of inaccurate
knowledge of environmental parameters.

\section{Conclusion \label{sec:concl}}

The main idea of the work is that, under conditions of inaccurate knowledge of
the environmental model, instead of the coefficient $K$ quantitatively
characterizing the proximity of the complex amplitudes of the calculated and
measured fields, the use of a coefficient $Q$ characterizing the proximity of
the intensity distributions of these fields in the phase space, can be more
effective. With the example of highly idealized waveguide model it is shown
that the coefficient $Q$, indeed, much weaker than $K$ varies with variations
in the environmental parameters and source position.

In Ref. \cite{V2017} some considerations are given explaining the stability of
field components formed by coherent states associated with small areas of the
phase space. These considerations can be used to justify the stability of the
intensity distribution $\left\vert a_{\mu}\right\vert ^{2}$ in phase space.

Another argument is that the ray line in the neighborhood of which the
insonified zone is localized is independent of frequency and it much less
"feels" the variations in the medium parameters than the complex field
amplitude. Therefore, it is natural to expect that the coefficient $Q$ is not
only much more stable than $K$ with respect to the inaccuracy of the medium
model, but it can be used at high frequencies, where the applicability of the
MFP is usually violated.

It should be understood, however, that the weak sensitivity of the coefficient
$Q$ to variations in the medium has its downside. In solving inverse problems,
the coefficient $Q$ will react weakly to changes in those unknown quantities
that must be recovered. This fact limits the ultimate accuracy of restoring an
unknown parameter. An example is the problem of source localization,
considered in Sec. \ref{sec:localization}. The main peak of uncertainty function
$K(r_{s},z_{s})$ is substantially narrower than the peak of function
$Q(r_{s\text{,}}z_{s})$. Therefore, under conditions of accurate knowledge of
the environment model, the traditional MFP method will allow to determine the
source position much more accurately. However, with an inaccurate environment
model, the MFP fails, while the MSP method allows a rough estimate of the
source position.

It is clear that the use of the similarity coefficient $Q$ is just one of the
possible variants of comparing expansions in the coherent states of the
measured and calculated fields. The analysis of other options is beyond the
scope of this paper.

Note that the practical use of the MSP method requires solving a number of
issues that were not addressed in this paper. In particular, these are
questions of the optimal choice of scale $\Delta_{z}$ and distance $d$, which
determines the boundary between the insonified and shadow zones. For the
applicability of the method, it is obviously required that the shadow zone
occupy a significant part of the phase plane area corresponding to admissible
depths and arrival angles. When working with CW signals, this condition can 
be satisfied only at short enough ranges. Here we did not consider the use of 
MSP for pulse signals, for which the applicability range of the method may become
wider due to the fact that the phase space acquires one more dimension (time).

\section*{Acknowledgment}

The author is grateful to Dr. L.Ya. Lyubavin and Dr. A.Yu. Kazarova for
valuable discussions. The work was supported by grants No. 15-02-04042 and
15-42-02390 from the Russian Foundation for Basic Research.

\end{document}